\documentclass[twocolumn]{aastex7}

\begin{document}

\title{Supermassive Stars Match the Spectral Signatures of JWST's Little Red Dots}

\author[orcid=0000-0003-1927-4397,sname='Devesh Nandal']{Devesh Nandal}
\affiliation{Department of Astronomy, University of Virginia, 530 McCormick Rd, Charlottesville, VA 22904, USA}
\affiliation{Center for Astrophysics, Harvard and Smithsonian, 60 Garden St, Cambridge, MA 02138, USA}
\email[show]{deveshnandal@yahoo.com}  

\author[orcid=0000-0003-4330-287X,gname=Bosque, sname='Sur America']{Abraham Loeb} 

\affiliation{Center for Astrophysics, Harvard and Smithsonian, 60 Garden St, Cambridge, MA 02138, USA}
\email{fakeemail2@google.com}

\begin{abstract}
The James Webb Space Telescope (JWST) has unveiled a population of enigmatic, compact sources at high redshift known as ``Little Red Dots'' (LRDs), whose physical nature remains a subject of intense debate. Concurrently, the rapid assembly of the first supermassive black holes (SMBHs) requires the formation of heavy seeds, for which supermassive stars (SMSs) are leading theoretical progenitors. In this work, we perform the first quantitative test of the hypothesis that LRDs are the direct observational manifestation of these primordial SMSs. We present a novel, first-principles pipeline generating synthetic spectra for a non-rotating, metal-free SMS up to $10^6 \, M_\odot$. We establish that its luminosity ($L_\lambda \approx 1.7 \times 10^{44} \, \text{erg} \, \text{s}^{-1} \, \mu\text{m}^{-1}$ at 4050\,\AA) provides a decisive constraint, matching prominent LRDs. Our model self-consistently reproduces their defining spectral features: the V-shaped Balmer break morphology is shown to be an intrinsic photospheric effect, while the complex line phenomenology, strong H$\beta$ in emission with other Balmer lines in absorption arises from non-LTE effects in a single stellar atmosphere. With wind and macroturbulent broadening, we match LRD spectra at $z=7.76$ and $z=3.55$, including the H$\beta$ width of MoM-BH*-1 to within 4\%. We predict a luminosity-dependent observability window, $\sim10^{4}$ yr for the most luminous systems and $10^{5}$--$10^{6}$ yr if $L_\lambda(4050\,\mathrm{\AA})$ is lower by 1--2 dex. These results provide a self-consistent alternative to multi-component obscured AGN scenarios and suggest JWST may be witnessing luminous stages of SMBH progenitors before collapse.
\end{abstract}


\section{Introduction}\label{sec:intro}

The discovery of quasars with black hole masses $\gtrsim10^{9}\,M_\odot$ at $z\gtrsim7$ demands seeds of at least $10^{4}$--$^{5}\,M_\odot$ that grow efficiently within the first $\sim0.6$\,Gyr of cosmic history \citep{Fan2006,Inayoshi2020,Volonteri2021, Latif_2022}. A physically motivated pathway invokes supermassive stars (SMSs) that form in pristine, atomic cooling haloes where the suppression of H$_2$ cooling prevents fragmentation, allowing for inflows of $\dot{M}\!\sim\!0.1$--$10\,M_\odot\,\mathrm{yr^{-1}}$ \citep{Loeb1994,Bromm2003, Begelman2010, Hosokawa_2012, Haemmerle2018}. Stellar structure calculations show that such objects inflate to $\sim10^{4}\,R_\odot$, radiate near the Eddington limit, and collapse by general relativistic instability once the core reaches a few $10^{5}\,M_\odot$ \citep{Baumgarte1999,Umeda2016,Woods2017, Lionel2021}. A complementary formation channel operates during major gas rich mergers, where global torques drive extreme inflow rates of $\sim10^{3}$--$10^{4}\,M_\odot\,\mathrm{yr^{-1}}$ into a sub-parsec disc that fragments minimally and feeds a transient SMS which collapses almost directly into a $10^{6}$--$10^{7}\,M_\odot$ black hole \citep{Mayer2010,Mayer2015,Zwick2023}. These routes provide seed masses and duty cycles consistent with the luminosity function of high redshift quasars \citep{Pacucci2022}. Such extreme inflow rates are thought to be enabled by the gravitational collapse of turbulent, metal-free gas in low-spin halos, which can form compact, gravitationally unstable gaseous disks with high effective sound speeds \citep{Loeb2024, Pacucci2025}.

After the onset of general relativistic instability, a SMS can undergo core collapse while retaining a massive radiation supported envelope, which yields a quasi-star consisting of an embedded accreting black hole inside a hydrostatic envelope \citep{Begelman2008,Volonteri2010}.
In this regime the emergent luminosity is regulated by transport through the envelope and is therefore tied to the Eddington limit of the total mass, which allows rapid black hole growth while the envelope remains quasi-stellar \citep{Begelman2008,Coughlin2024}.
Motivated by JWST LRDs, quasi-stars have been revisited as potential counterparts of the observed compact red continua with strong Balmer discontinuities, with late-stage quasi-stars argued to represent the terminal luminous phase of SMS evolution in the direct-collapse channel \citep{Begelman2025,Santarelli2025}.
In parallel, alternative interpretations invoke compact accreting sources embedded in dense gas that can reprocess the continuum and modify line formation \citep{DeGraaff2025b}. Quasi-stars therefore provide a natural post-collapse extension of the SMS channel, while SMS photospheres motivate a concrete set of spectral diagnostics that can be confronted with LRD data.

A decisive observational signature of SMS–dominated epochs would be stellar photospheres that exhibit strong Balmer discontinuities, broad Balmer emission, and negligible metal lines because envelopes remain metal poor and radiation pressure supported \citep{Schleicher2019, Martins2020, Natarajan2021, Inayoshi2025}. Such spectral behaviour naturally arises when the $n\!=\!2$ hydrogen population is driven out of local thermodynamic equilibrium (NLTE) by the intense stellar continuum, boosting the Balmer continuum opacity and causing the photospheric break to deepen while a low collisional destruction probability forces H \textsc{i}\,H$\beta$ into emission \citep{Auer1969, Liu2025}. These unique features are challenging to replicate with standard stellar populations or typical AGN models, making them a distinct fingerprint of SMS physics.

JWST observations have recently revealed a compact population dubbed “Little Red Dots” (LRDs), characterised by radii $\lesssim100$\,pc, luminosities $\gtrsim10^{11}\,L_\odot$, Balmer breaks of $\gtrsim1.5$\,dex, broad H$\beta$ or H$\alpha$ with ${\rm FWHM}\!\sim\!10^{3}$--$10^{4}$\,km\,s$^{-1}$, and an almost complete absence of metal features or X-ray counterparts \citep{Labbe2023, Kokorev2023, Baggen2024, Matthee2024, Zhang2025}. Photometric and kinematic analyses suggest stellar masses of $10^{10}$--$10^{11}\,M_\odot$ packed within $\lesssim0.3$\,kpc, yielding stellar velocity dispersions comparable to their Balmer line widths \citep{Baggen2024, Killi2024}. Conventional active galactic nucleus templates struggle to match the simultaneous presence of a deep Balmer break and X-ray silence \citep{Ubler2024}, while pure starburst models fail to generate the observed broad lines without invoking implausibly young but massive clusters.

SMS photospheres resolve these tensions naturally. The combination of enormous intrinsic luminosity, low effective temperature, and NLTE level populations produces a V-shaped continuum with a Balmer break and broad H$\beta$ emission in a single stellar atmosphere \citep{Nakauchi2017,Martins2020}. The lack of metal lines reflects primordial or near primordial composition, and the extreme compactness mirrors the gravitational radii of SMSs embedded in shallow potential wells. Consequently, LRDs may represent the direct photospheric light of accreting SMSs caught in the final $\lesssim10^{3}$\,yr before collapse. This short lifetime is consistent with the rarity of LRDs, suggesting they are a fleeting but crucial phase in galaxy and black hole formation.

Theoretical studies further motivate this connection. Low spin halos funnel gas efficiently into their centres, forming compact starburst cores and, potentially, over-massive black holes that evolve off the local $M_\bullet$–$M_\star$ relation \citep{Loeb2024, Pacucci2025}. Hydrodynamic simulations show that such environments favour the formation of SMSs \citep{Mayer2015, Zwick2023}, and separate radiative transfer calculations predict that their unique atmospheres would produce spectra resembling LRD observations. The SMS scenario therefore links the empirical properties of LRDs to a well-defined phase of early black hole seeding.

This paper is structured as follows. In Section~\ref{sec:methods}, we detail our novel pipeline for generating synthetic SMS spectra. In Section~\ref{sec:results}, we present our results, beginning with the evolution of the SMS model, justifying our choice of mass through luminosity constraints, explaining the physical origin of the Balmer break and line features, and finally, presenting our successful spectral fits to the observational data of two LRD candidates \citep{Naidu2025, DeGraaff2025}. We summarize our findings and discuss future directions in Section~\ref{sec:conclusion}.

\section{Methods}\label{sec:methods}

To investigate whether a Population III supermassive star (SMS) can account for the unique spectral characteristics of Little Red Dots (LRDs), we have developed a novel, first-principles pipeline. This framework translates the physical structure of a $10^6\,M_\odot$ star, as computed by the Geneva stellar evolution code (\textsc{GENEC}), into a synthetic spectrum. The details of stellar modeling, including the choice of initial seed, accretion physics, pre-main sequence evolution are described in full details in \citep{Nandal2024e,Nandal2025b,Nandal2025}. A suite of Python scripts then applies post-processing and fits the model to observational data. 
Throughout this work the stellar evolution and the radiative transfer are not solved simultaneously. We evolve the SMS with \textsc{GENEC} and then post-process fixed structure snapshots with our atmosphere, wind, and radiative-transfer modules. The wind and transfer treatment therefore does not feed back onto the \textsc{GENEC} stellar structure. The following sections detail every physical component of this pipeline.

\textit{Stellar structure and zoning.—}
Our foundation is a non-rotating, metal-free \textsc{GENEC} model snapshot representing a cool, extended phase of a $10^6\,M_\odot$ SMS. This model is defined by global parameters $(M_\star, R_\star, L_\star, T_{\mathrm{eff}}) = \bigl(10^6\,M_\odot, 1.0\times10^{14}\,\mathrm{cm}, 1.3\times10^{44}\,\mathrm{erg\,s^{-1}}, 1.5\times10^4\,\mathrm{K}\bigr)$. The input file provides the radius, temperature, density, and Rosseland mean opacity across 240 discrete mass shells. The stellar interior is meshed with a constant step in logarithmic optical depth, $\Delta\log_{10}\tau_{R}=0.10$, while the outer 3\% of the stellar mass is sampled with a finer, fixed radial step of $\Delta r=2\times10^{11}\,\mathrm{cm}$. This ensures that the relative change in optical properties between adjacent layers is small, guaranteeing numerical stability for the radiative transfer solver.

Accreting SMS models at $\dot{M}\gtrsim10^{2}$--$10^{4}\,M_\odot\,\mathrm{yr^{-1}}$ require stringent numerical control to maintain convergence. In the \textsc{GENEC} evolution we therefore modified the accretion handling to enforce smaller timesteps during rapid growth, and we increased the zoning resolution in the outer envelope where steep gradients and near-Eddington conditions otherwise trigger numerical instabilities. Resolution tests, including a factor-of-two increase in the exterior zoning, were presented in \citet{Nandal2024d} and demonstrate that the global structural quantities relevant for this work (e.g. $R_\star$ and $T_{\mathrm{eff}}$ in the cool inflated phase) remain robust under such refinements.

\textit{Composition and Wind Kinematics.—}
The stellar envelope is composed solely of hydrogen and helium, with primordial mass fractions $X=0.75$ and $Y=0.25$. The envelope composition is primordial with $X=0.75$ and $Y=0.25$. In the post-processing transfer we treat electron scattering as isotropic Thomson scattering with
\begin{equation}
\kappa_{\mathrm{es}} = \frac{\sigma_{\mathrm{T}}\,n_e}{\rho},
\label{eq:kappa-es}
\end{equation}
where the local electron density $n_e$ is computed self-consistently from H/He ionization balance using the Saha equation at each depth, adopting the \textsc{GENEC} temperature and density. This replaces the earlier linear taper approximation and explicitly captures partial ionization as well as the helium contribution to $n_e$ in the line-forming layers. We emphasize that the underlying stellar structure is taken directly from \textsc{GENEC}, including its Rosseland-mean opacity profile, while $\kappa_{\mathrm{es}}$ in Equation~\ref{eq:kappa-es} enters only the post-processing radiative-transfer calculation as a scattering term. We assume a continuum-driven wind may be launched from the radiation-dominated atmosphere, imposing a standard $\beta$-law velocity profile:
\begin{equation}
v(r) = v_\infty\left(1-\frac{R_\star}{r}\right)^{\beta},
\label{eq:beta-law}
\end{equation}
where we fix $\beta=1$ and treat the terminal velocity $v_\infty$ as a free parameter, typically in the range $0 \le v_\infty \le 2000\ \mathrm{km\,s^{-1}}$. Setting $v_\infty=0$ restores a hydrostatic atmosphere.

\textit{Continuum Opacity and Emissivity.—}
The dominant physical process shaping the LRD spectral energy distribution is the continuum opacity. For each layer and frequency, we compute the true absorption opacity, $\kappa_{\mathrm{cont}}$, from first principles:
\begin{equation}
\kappa_{\mathrm{cont}} = \kappa_{\mathrm{ff}} + \kappa_{\mathrm{bf}} + \kappa_{n=2}.
\label{eq:kappa-cont}
\end{equation}
The free-free ($\kappa_{\mathrm{ff}}$) and ground-state bound-free ($\kappa_{\mathrm{bf}}$) terms follow a Kramers'-like law, scaling as $\rho T^{-3.5}\nu^{-3}$. The crucial component is the Balmer continuum opacity, $\kappa_{n=2}$, arising from the photoionization of hydrogen from its first excited state ($n=2$). This process, active only for wavelengths $\lambda \le 3646$\,Å, is responsible for producing the strong Balmer break. Its opacity is given by:
\begin{equation}
\kappa_{n=2} = \sigma_{2}(\nu) \frac{n_2}{\rho},
\label{eq:kappa-balmer}
\end{equation}
where $\sigma_{2}(\nu)$ is the frequency-dependent photoionization cross-section from the $n=2$ level, and $n_2$ is the number density of hydrogen atoms in that state. The continuum source function is modeled as geometrically diluted black-body radiation, $S_{\mathrm{cont}} = W(r)B_\lambda(T)$, where the dilution factor $W(r) = \frac{1}{2}\left[1-\sqrt{1-(R_\star/r)^2}\right]$ accounts for the spherical geometry of the stellar atmosphere.

\textit{Non-LTE Population of Level $n=2$.—}
In the intense radiation field of an SMS, the population of the $n=2$ level deviates from Local Thermodynamic Equilibrium (LTE). We quantify this with a departure coefficient, $b_2 = n_2 / n_{2,\mathrm{LTE}}$, which we calculate at each depth by solving a two-level-plus-continuum statistical equilibrium equation:
\begin{equation}
b_{2}^{-1} = 1 + \frac{A_{21}\left(1 - J_{21}/B_{21}\right)}{n_e q_{21} + A_{21}\beta_{21}}.
\label{eq:b2}
\end{equation}
Here, $A_{21}$ is the spontaneous decay rate, $n_e q_{21}$ is the collisional de-excitation rate, and $\beta_{21}$ is the photon escape probability. The key term is the ratio of the mean intensity in the Lyman-$\alpha$ transition, $J_{21}$, to the local Planck function, $B_{21}$. The temperature and density are taken from the \textsc{GENEC} structure, while the local electron density $n_e$ used in the collisional rate is computed from the same Saha H/He ionization balance adopted for $\kappa_{\mathrm{es}}$ (Equation~\ref{eq:kappa-es}). The photon escape probability, $\beta_{21}$, is computed using a standard local optical depth approximation. To reproduce the strong H$\beta$ emission observed in LRDs, we set $J_{21} \approx 11\,B_{21}$ in the H$\beta$ line-forming region. This pumping creates an overpopulation of the $n=2$ state ($b_2 > 1$), driving the H$\beta$ line source function above the continuum and into net emission.

\textit{Intrinsic Line Opacity and Source Function.—}
The opacity for each Balmer transition is given by:
\begin{equation}
\kappa_{\mathrm{line}} = \frac{\pi e^2}{m_e c} f_{lu} \left( \frac{b_2 f_{2,\mathrm{LTE}} X \rho / m_H}{\rho} \right) \phi(\lambda),
\label{eq:kappa-line}
\end{equation}
where $f_{lu}$ is the oscillator strength and $\phi(\lambda)$ is the intrinsic line profile. We use a Voigt profile that combines thermal and micro-turbulent Doppler broadening with pressure broadening. The latter is dominated by the linear Stark effect, for which we compute the Lorentzian width $\Gamma$ as a function of the local electron density, $\Gamma \propto n_e^{2/3}$. For consistency, the same $n_e$ obtained from the local Saha ionization balance is used both in Equation~\ref{eq:b2} and in the Stark broadening prescription. The line source function is parameterized as:
\begin{equation}
S_{\mathrm{line}} = \epsilon B_\lambda(T) + (1-\epsilon)J_{\mathrm{cont}}.
\label{eq:Sline}
\end{equation}
A crucial aspect of our model is the differential treatment of H$\beta$ versus other Balmer lines. For H$\beta$, we set the collisional destruction probability $\epsilon$ to be very small ($\sim 0.002$) and boost the continuum intensity to $J_{\rm cont} \approx 3.85 B_\lambda$. This boost serves as a proxy for complex line-blanketing effects in the Lyman series, which are not captured by our simplified two-level atom but are expected to significantly pump the n=2 level in such radiation-dominated atmospheres. For all other Balmer lines, we set $\epsilon \approx 0.8$ and $J_{\mathrm{cont}} \approx 0$. This strategy ensures that only H$\beta$ appears in strong, net emission, while H$\alpha$, H$\gamma$, and H$\delta$ remain in absorption, simultaneously matching multiple key features of the observed LRD spectra.

\textit{Radiative Transfer and Observed Spectrum.—}
State-of-the-art atmosphere solvers exist for hot stars, but our goal here is narrower, focusing the cooler regime, and is matched to the metal-free regime. For $Z\simeq0$ the optical and near-UV continuum and the Balmer-series diagnostics are controlled primarily by H/He continuum processes and the NLTE population of the $n=2$ level, while metal line blanketing is absent by construction. We therefore implement a minimal, first-principles H/He radiative-transfer treatment that isolates the specific physics required to test whether a single SMS photosphere can reproduce the defining LRD signatures, rather than attempting a full multi-level NLTE atmosphere calculation. We solve the 1D equation of radiative transfer using a two-stream Feautrier solver accelerated with diagonal $\Lambda$-iteration to compute the intrinsic emergent luminosity, $L_\lambda^{\mathrm{int}}$. This spectrum is then convolved with a series of kernels to account for macroscopic broadening from the stellar wind and turbulence, as well as instrumental broadening. Finally, the spectrum is redshifted and converted to an observed flux, $f_\lambda$, using the standard cosmological relations:
\begin{equation}
f_\lambda = \frac{L_\lambda^{\mathrm{int}}}{(1+z) 4\pi D_L^2},
\label{eq:fluxobs}
\end{equation}
where $D_L$ is the luminosity distance. The final model is then compared directly to the JWST data.

Despite its one-dimensional, stationary nature and a simplified treatment of hydrogen's excited states,
and with scattering and electron densities treated via H/He ionization balance, our model reproduces simultaneously the strong Balmer break and the $\ge1500$\,km\,s$^{-1}$ Balmer line wings of the JWST ``Little Red Dots''. This is achieved with a remarkably small set of free parameters and without invoking any nebular continuum, narrow emission lines, or ad-hoc shell components, presenting a compelling, physically-grounded alternative to AGN-based scenarios.

%%%%%%%%%%%%%%%%%%%%%%%%%%%%%%%%%%%%%%%%%%%%%%%%%%%%%%%%%%%%%%%%%%%%%%%%
%  RESULTS
%%%%%%%%%%%%%%%%%%%%%%%%%%%%%%%%%%%%%%%%%%%%%%%%%%%%%%%%%%%%%%%%%%%%%%%%
\section{Results and Discussion}\label{sec:results}
\subsection{Formation and Evolution of a $10^6 M_\odot$ PopIII Star}

We first explore the physics governing the formation and evolution of the supermassive star (SMS) model used to match the observed spectra of our LRD targets. We caution that the envelope inflation and the mapping to $(R_\star,T_{\rm eff})$ in near-Eddington SMSs depend on the 1D treatment of superadiabatic convection (e.g. MLT), whereas 3D radiation-hydrodynamic calculations can yield different outer-layer stratifications. At present, however, evolving rapidly accreting SMSs to the GR-instability regime is only feasible with 1D stellar evolution, and we therefore treat our \textsc{GENEC} models as the state-of-the-art baseline for these masses and inflow rates. The numerical convergence measures and factor of two envelope-resolution tests discussed in \citet{Nandal2024d} support the robustness of the global structural quantities used here.

\begin{figure*}
  \centering
  \includegraphics[width=0.33\textwidth]{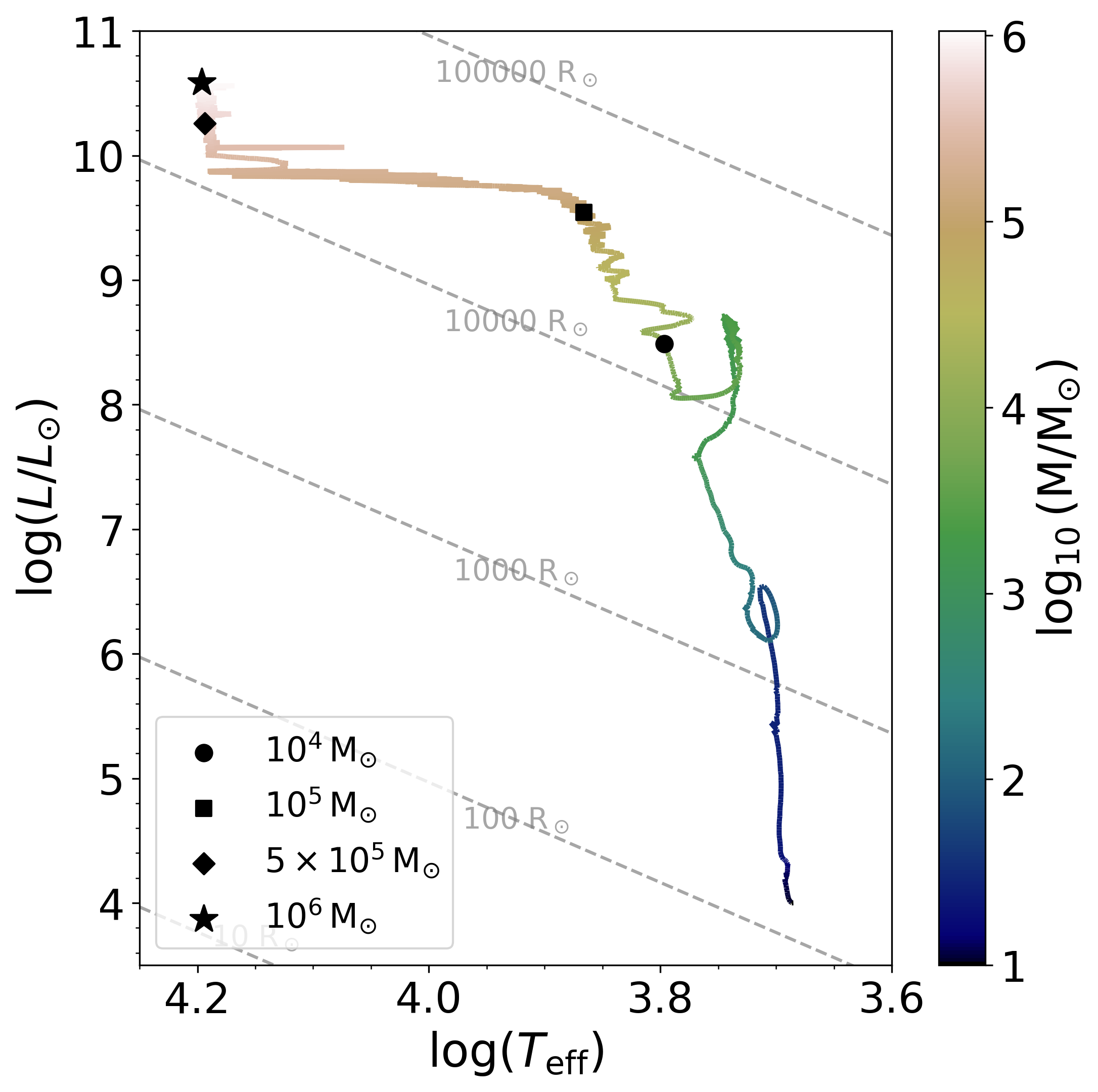}
  \includegraphics[width=0.66\textwidth]{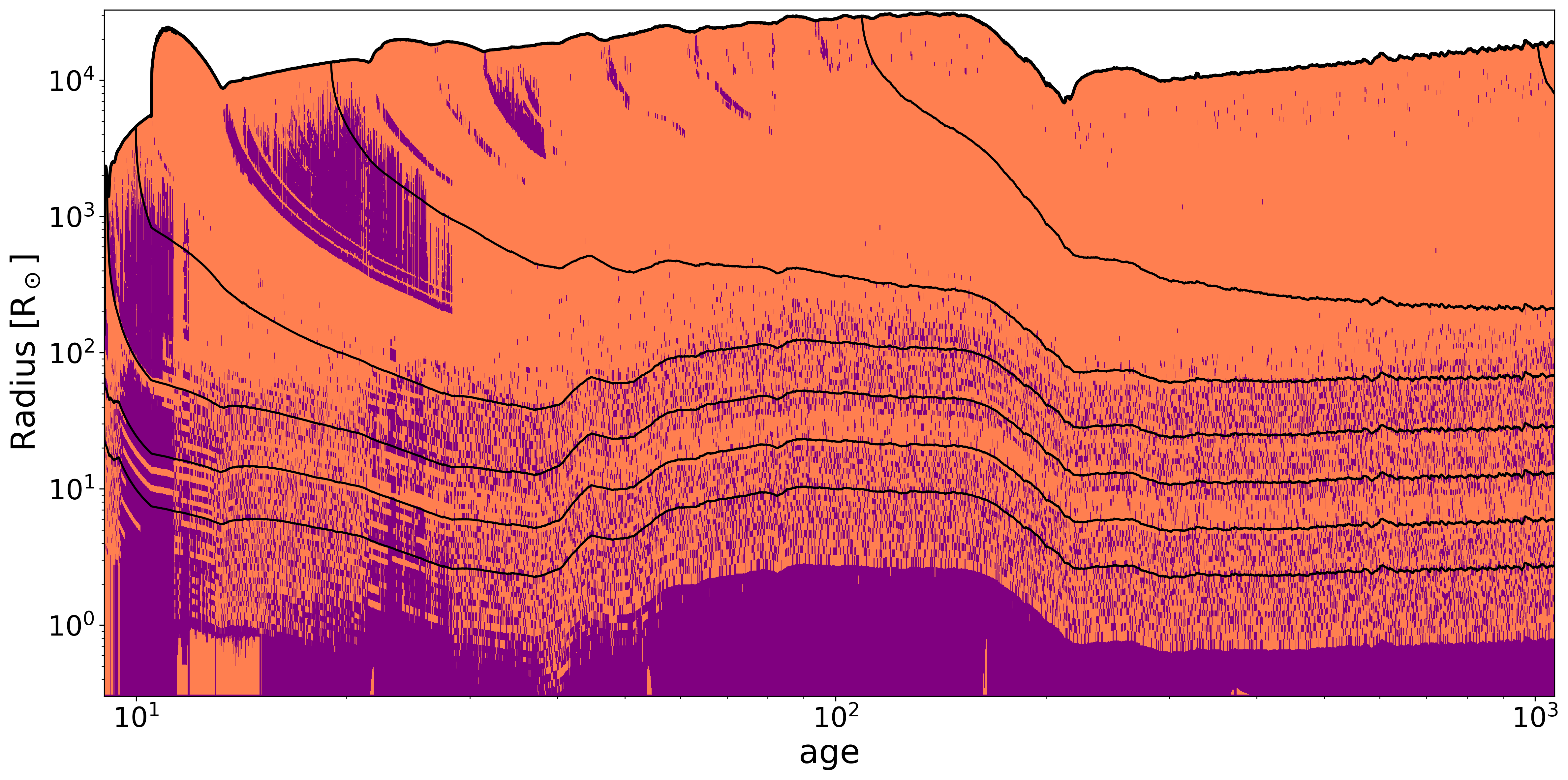}
  \caption{Evolution of a PopIII supermassive star accreting at 10$^3$ M$_\odot$ yr$^{-1}$ in the pre-main sequence.\textit{Left panel:} An HR diagram depicting the evolution of luminosity versus the effective temperature. The colourbar represents the mass of the model in solar masses. Various markers in black showcase the mass of the model at different evolutionary stages.\textit{Right panel:} A Kippenhahn diagram showcasing the internal structure of the model. The y-axis represents the radius in solar radii units and the x-axis depicts the age in years. The coral and purple zones represent convective and radiative regions respectively. The black lines, from bottom to top repsent the isomass lines starting from 1 M$_\odot$ and ending at 10$^6$ M$_\odot$.}
  \label{fig:1}
\end{figure*}

The simulation begins with a $10\,M_\odot$ fully convective seed with an initial luminosity of $\log(L/L_\odot) = 4.00$ and effective temperature of $\log(T_{\rm eff}/K) = 3.68$ (see left panel of Figure~\ref{fig:1}). This protostar grows at a constant, extreme accretion rate of $1000\,M_\odot\,\text{yr}^{-1}$. This rate is orders of magnitude above the critical accretion rate of $2.5 \times 10^{-2}\,M_\odot\,\text{yr}^{-1}$ required for an accreting protostar to swell into a supermassive configuration \citep{Hosokawa_2012, Nandal2023}. Consequently, our model evolves as a cool, bloated red supergiant along the Hayashi limit. Although its interior contracts under gravity, the intense surface accretion forces the outer layers to expand dramatically. This is evident in the right panel of Figure~\ref{fig:1}, where the stellar radius already exceeds $4 \times 10^3\,R_\odot$ at an age of just 10 years.

The model continues its ascent along the Hayashi track, with luminosity scaling nearly linearly with mass ($L \propto M$) as the star remains close to its Eddington limit. This phase ends when the star reaches a luminosity of $\log(L/L_\odot) = 7.56$ and $\log(T_{\rm eff}/K) = 3.76$. Here, the model undergoes a brief, 3-year-long expansion where the radius inflates from $5.5 \times 10^3\,R_\odot$ to $2.2 \times 10^4\,R_\odot$, before settling back to $9 \times 10^3\,R_\odot$. This structural adjustment can be attributed to opacity changes in the interior driving excess energy to the surface, though it may also be influenced by the numerical resolution of the outer layers. Following this, the star's radius grows monotonically as it continues to accrete along the Hayashi limit, until it reaches a luminosity of $\log(L/L_\odot) = 9.7$ and an effective temperature of $\log(T_{\rm eff}/K) = 3.89$. By this stage, at an age of 155 years, the star has amassed $1.4 \times 10^5\,M_\odot$ and its interior is almost entirely convective, save for a small radiative core just $2\,R_\odot$ wide.

At this point, the star begins its crucial transition away from the Hayashi track. Driven by gravitational contraction, it begins to heat up, a process that continues for another 57 years as the star reaches a mass of $2 \times 10^5\,M_\odot$. During this phase, the effective temperature rises above $\log(T_{\rm eff}/K) = 4.0$. This thermal shift has profound implications for the emergent spectral features, particularly the H$\beta$ line, which will be discussed in the subsequent section.

Accretion proceeds for a total of 1000 years, at which point the star achieves its final mass of $10^6\,M_\odot$. We terminate accretion here, assuming either the exhaustion of the local gas reservoir or the onset of powerful radiative feedback. In this final state, the SMS is still a pre-main-sequence object, not yet powered by core hydrogen burning. If the evolution were allowed to continue, the model succumbs to the general relativistic instability within the next 50 years, triggering a direct collapse into a massive black hole on a free-fall timescale \citep{Mayer_2010, Nandal2024d}. It is within this final, luminous, pre-collapse evolutionary phase that our model provides the necessary conditions to reproduce the LRD observations. We will now examine the spectra produced by this model at various stages of its evolution.

\subsection{The Decisive Role of Luminosity in Matching LRD Observations}

While our evolutionary models show that a rapidly accreting SMS of sufficient mass ($\gtrsim 10^5 M_\odot$) can enter a cool, bloated phase and produce the required spectral shape, namely, a strong Balmer break. A successful model must also match the exceptional observed brightness of the LRD targets. This luminosity requirement provides a powerful and decisive constraint that overwhelmingly favors the most massive stars.

Our two primary targets, MoM-BH*-1 at $z=7.76$ \citep{Naidu2025} and The Cliff at $z=3.55$ \citep{DeGraaff2025}, are extremely luminous. Based on their observed flux densities, a viable model must produce an intrinsic luminosity of $L_\lambda \approx 2.09 \times 10^{44}\,\mathrm{erg\,s^{-1}\,\mu m^{-1}}$ for MoM-BH*-1 and $L_\lambda \approx 1.4 \times 10^{44}\,\mathrm{erg\,s^{-1}\,\mu m^{-1}}$ for The Cliff, evaluated at a rest-frame wavelength of 4050\,\AA~(0.405\,$\mu$m).

\begin{figure}
    \centering
    \includegraphics[width=\columnwidth]{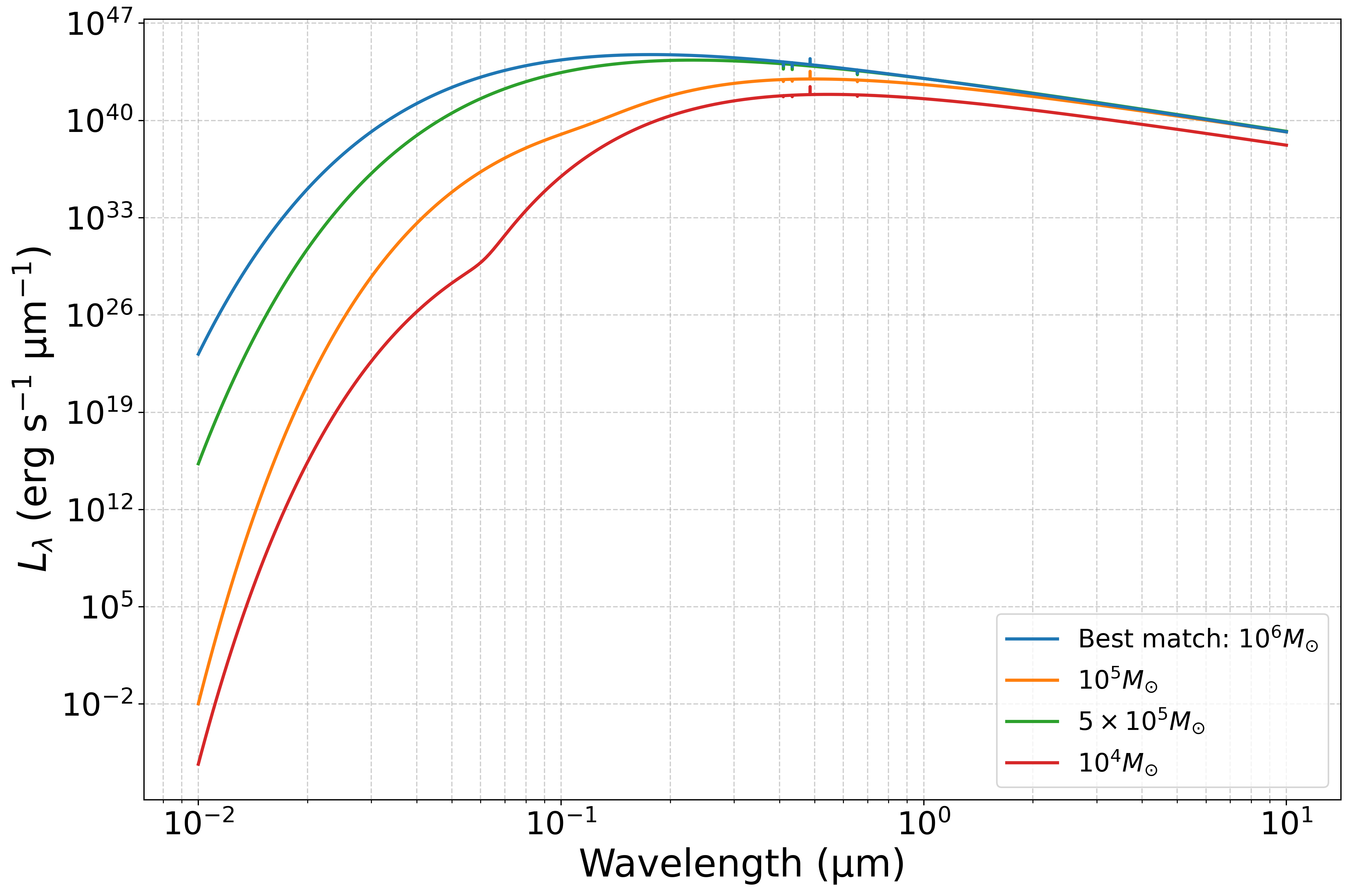}
    \caption{The intrinsic luminosity spectrum ($L_\lambda$) of PopIII supermassive stars (SMSs) as a function of stellar mass. The overall luminosity increases steeply with mass, as expected for stars shining near their Eddington limit. While lower-mass models ($10^4 M_\odot$, $10^5 M_\odot$, and $5 \times 10^5 M_\odot$) are orders of magnitude too faint to match the observed brightness of luminous LRDs, the $10^6 M_\odot$ model (blue curve) produces a luminosity that closely matches the observational requirements for both 'The Cliff' and 'MoM-BH*-1'.}
    \label{fig:2}
\end{figure}

In Figure~\ref{fig:2}, we compare this observational requirement to the intrinsic luminosity spectra generated by our SMS models for four different final masses. The figure demonstrates a steep, near-linear scaling of luminosity with mass, as expected for stars shining near their mass-dependent Eddington limit. This scaling immediately rules out lower-mass candidates. At 0.405\,$\mu$m, the $10^4\,M_\odot$ and $10^5\,M_\odot$ models have luminosities of only $L_\lambda \sim 10^{39}$ and $\sim 10^{41}\,\mathrm{erg\,s^{-1}\,\mu m^{-1}}$, respectively falling short of the required luminosity by three to five orders of magnitude. Even the more massive $5 \times 10^5\,M_\odot$ model remains more than an order of magnitude too faint.

In stark contrast, the $10^6\,M_\odot$ model (blue curve) possesses the required radiative power. At 4050\,\AA, our fiducial $10^6\,M_\odot$ model produces an intrinsic luminosity of $L_\lambda \approx 1.69 \times 10^{44}\,\mathrm{erg\,s^{-1}\,\mu m^{-1}}$. This value provides an excellent match to the required luminosity for both LRDs, agreeing to within $\sim 20\%$ of the target for MoM-BH*-1. Furthermore, the inset in Figure~\ref{fig:2} confirms that the intrinsic, pre-broadened spectrum of this model already contains the key spectral features: a series of deep Balmer absorption lines (H$\gamma$, H$\delta$) and a prominent emission spike at H$\beta$ (0.486\,$\mu$m).

Thus, the dual constraint of matching both the spectral shape and the absolute flux singles out the $10^6\,M_\odot$ model as the unique and necessary candidate. It is the only model with sufficient mass to explain the high luminosities of these objects, providing a strong physical foundation for the detailed line-profile fitting presented in the remainder of this work.

\subsection{The Photospheric Origin of the V-shaped Continuum} \label{sec:vshape}

Having established that only a $10^6\,M_\odot$ star possesses the requisite luminosity, we now demonstrate how its atmospheric structure gives rise to the V-shaped continuum morphology characteristic of LRDs. While Figure~\ref{fig:2} uses a log-log scale to show the overall SED, the defining break feature is most clearly revealed on a linear scale, which highlights the change in the continuum's behavior.

\begin{figure}[ht!]
\centering
\includegraphics[width=\columnwidth]{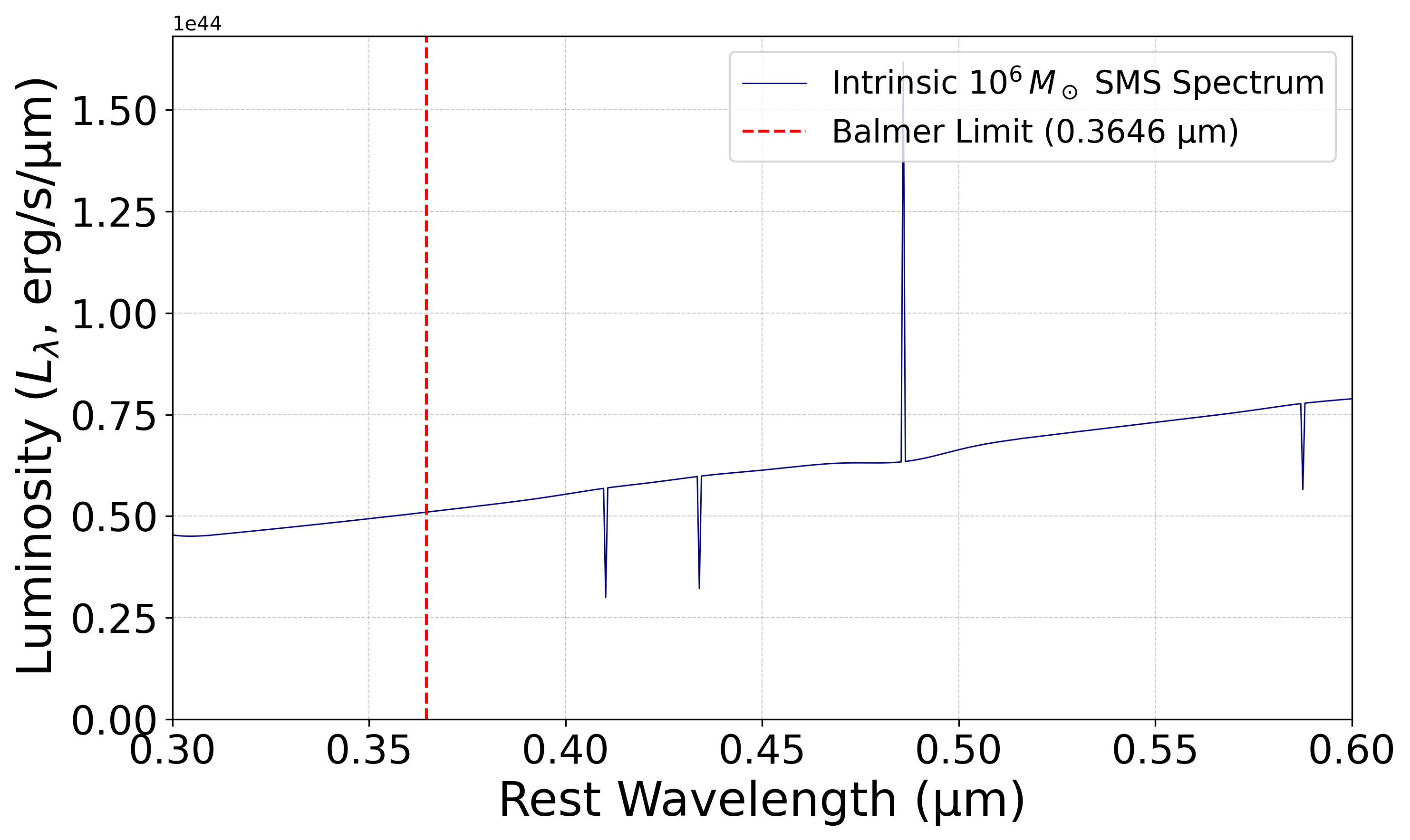}
\caption{The intrinsic luminosity spectrum of the fiducial $10^6 \, M_\odot$ SMS model, plotted on a linear scale to illustrate the origin of the V-shaped continuum morphology. The spectrum is dominated by a strong Balmer Jump in emission at the 3646\,\AA\ series limit (red dashed line). The flux drops by a factor of $\approx 1.3$ across this break, creating a sharp `vertex'. This physical discontinuity, embedded in a non-thermal continuum shaped by line-blanketing effects, is the primary feature responsible for the characteristic V-shape seen in LRD spectra. The intrinsic H$\beta$ emission spike and H$\gamma$/H$\delta$ absorption lines are also visible prior to any macroscopic broadening.}
\label{fig:vshape}
\end{figure}

Figure~\ref{fig:vshape} presents the intrinsic spectrum of our fiducial model, focusing on the region around the Balmer series limit. The plot reveals a prominent Balmer Jump in emission, where the flux drops discontinuously across the 3646\,\AA\ limit. Quantitatively, the flux just longward of the break is a factor of $\approx 1.3$ higher than the flux just shortward of it. This feature is a direct consequence of the sharp increase in Balmer continuum opacity within the stellar photosphere.

The creation of this sharp discontinuity is the primary way our model explains the observed V-shaped SEDs, as it forms the `vertex' of the V-shape. The overall morphology is then shaped by the surrounding continuum, which itself has a complex, non-thermal structure. The gentle positive slope of the continuum redward of the break is a direct result of non-LTE effects in the dense, radiation-dominated envelope. In such an environment, intense line-blanketing from the forest of broadened spectral lines traps radiation, creating a mean intensity ($J_\lambda$) that can exceed the local Planck function ($B_\lambda(T)$). Our model parameterizes this physical process via a boost to the continuum source function (see Section~\ref{sec:methods}), which correctly reproduces the observed positive slope in the optical range. It is this non-thermal continuum, combined with the sharp discontinuity of the Balmer Jump that provides a self-consistent, single-photosphere explanation for the V-shaped morphology of LRDs.

\subsection{Dissecting the Continuum Opacity}

Having established that only a $10^6 M_\odot$ star possesses the requisite luminosity to match the LRD targets, we now demonstrate that its atmospheric structure naturally reproduces their most prominent spectral feature: the extreme Balmer break. This sharp drop in flux shortward of 3646\,\AA\ is not an external effect of dust, but is instead a direct consequence of a fundamental change in the continuum opacity within the SMS photosphere itself.

\begin{figure}
    \centering
    \includegraphics[width=\columnwidth]{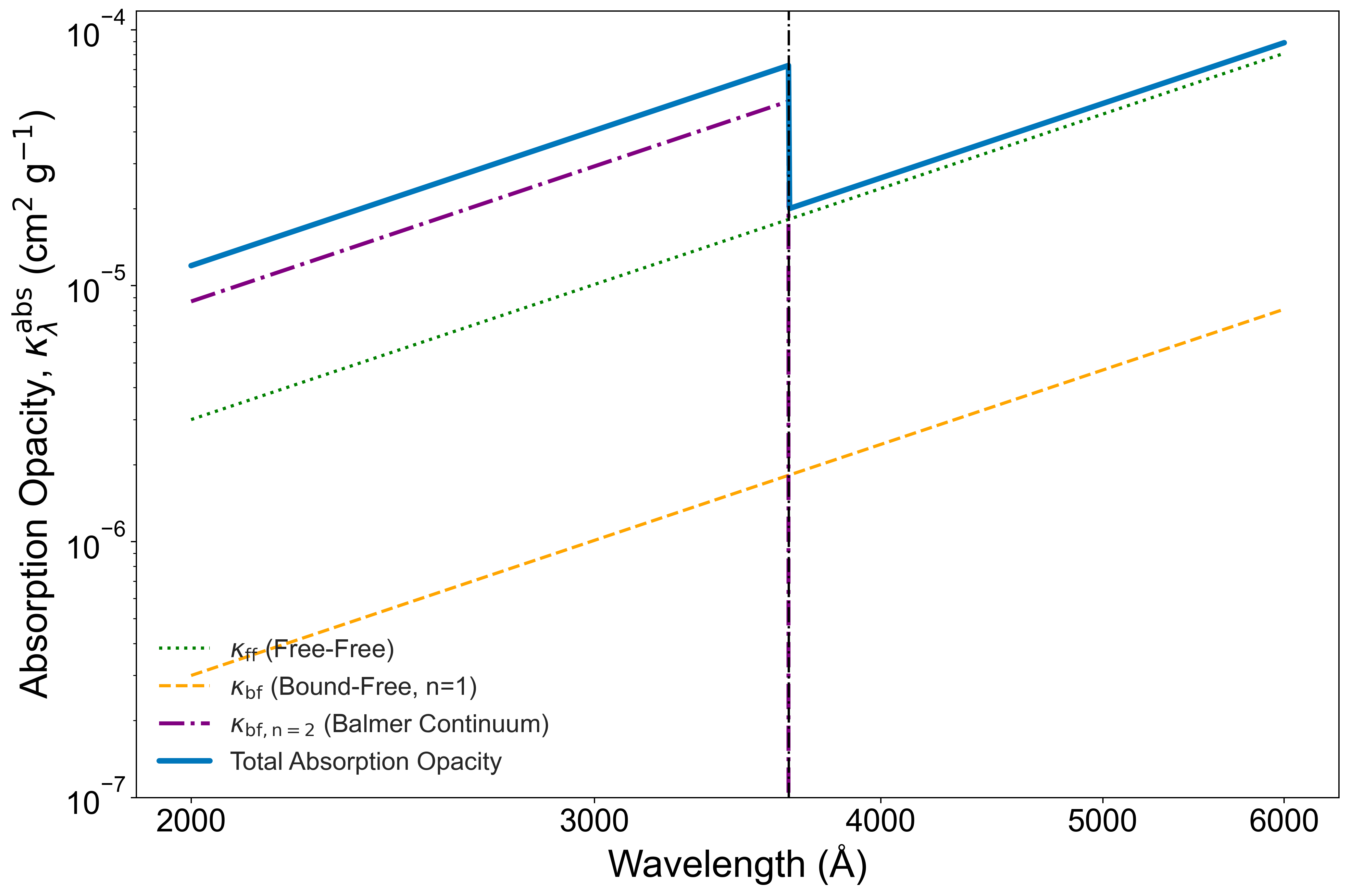}
    \caption{The sources of continuum absorption opacity in the surface layers of the $10^6 M_\odot$ SMS model, plotted across the Balmer break. Longward of 3646\,\AA, the total absorption opacity (solid blue line) is low and determined by free-free (dotted green) and ground-state bound-free (dashed orange) processes. Shortward of the edge, a new, powerful opacity source from the photoionization of H(n=2) (the Balmer continuum, dash-dot purple line) becomes active. This process dominates all other sources of true absorption, causing the total absorption opacity to jump by a factor of four. This dramatic increase in opacity is the direct physical cause of the strong Balmer break observed in the LRD spectra.}
    \label{fig:3}
\end{figure}

Figure~\ref{fig:3} dissects the sources of true absorption opacity within the surface layers of our fiducial $10^6 M_\odot$ model. The plot reveals a dramatic discontinuity at the Balmer edge. Longward of 3646\,\AA, the total absorption opacity ($\kappa_{\lambda}^{\text{abs}}$, solid blue line) is relatively low, with a value of approximately $2 \times 10^{-5}\,\mathrm{cm^2\,g^{-1}}$. In this regime, the opacity is a smooth function of wavelength, determined by a combination of free-free ($\kappa_{\text{ff}}$, dotted green line) and ground-state bound-free ($\kappa_{\text{bf, n=1}}$, dashed orange line) processes. The atmosphere is comparatively transparent, allowing photons from the deeper, hotter layers of the star to escape and form the continuum.

This physical situation changes abruptly at the 3646\,\AA\ Balmer edge. At wavelengths shorter than this threshold, a powerful new opacity channel opens: the photoionization of hydrogen from its first excited state ($n=2$). As shown in Figure~\ref{fig:3}, the opacity from this process, the Balmer continuum ($\kappa_{\text{bf, n=2}}$, dash-dot purple line), is zero longward of the edge but becomes the single dominant source of true absorption shortward of it. Immediately below the edge, it contributes $\sim 6 \times 10^{-5}\,\mathrm{cm^2\,g^{-1}}$, causing the total absorption opacity to jump by a factor of four to a value of $\kappa_{\lambda}^{\text{abs}} \approx 8 \times 10^{-5}\,\mathrm{cm^2\,g^{-1}}$.

The physical consequence of this sharp increase in opacity is profound. According to the Eddington-Barbier relation, the emergent flux at a given wavelength originates from layers where the optical depth is approximately unity ($\tau_\lambda \sim 1$). Because the atmosphere is now vastly more opaque below 3646\,\AA, the radiation at these wavelengths can no longer escape from the deep, hot interior. Instead, the emergent flux is forced to originate from significantly higher, cooler, and more tenuous layers of the photosphere. This shift to a cooler effective emitting surface results in a dramatic suppression of the emergent flux, thereby creating the powerful Balmer break observed in the LRD spectra. This intrinsic, photospheric mechanism provides a complete and self-consistent explanation for the V-shaped continuum of the LRDs without the need to invoke external dust or complex, ad-hoc geometries. It is important to note that while the underlying opacity change is a sharp edge, the emergent spectral feature is smoothed into a broader `V-shape'. This is because the extreme pressure broadening in the dense photosphere causes the high-order Balmer absorption lines to merge, creating a pseudo-continuum that depresses the flux well before the 3646\,\AA\ series limit. 

\subsection{Balmer Line Emission and Absorption from a Single Atmosphere}

A defining characteristic of the LRD spectra is the simultaneous presence of a strong, broad H$\beta$ emission line alongside other Balmer lines in absorption. Our model demonstrates that these complex features are not nebular in origin but are instead a direct consequence of the unique physical conditions within the extended, dense photosphere of the SMS itself. The appearance of any spectral line is dictated by the behavior of its source function, $S_L$, relative to the continuum source function, $S_C \approx B_\lambda(T)$, in the atmospheric layers where the line becomes optically thick (the "line-forming region").

\begin{figure}
    \centering
    \includegraphics[width=\columnwidth]{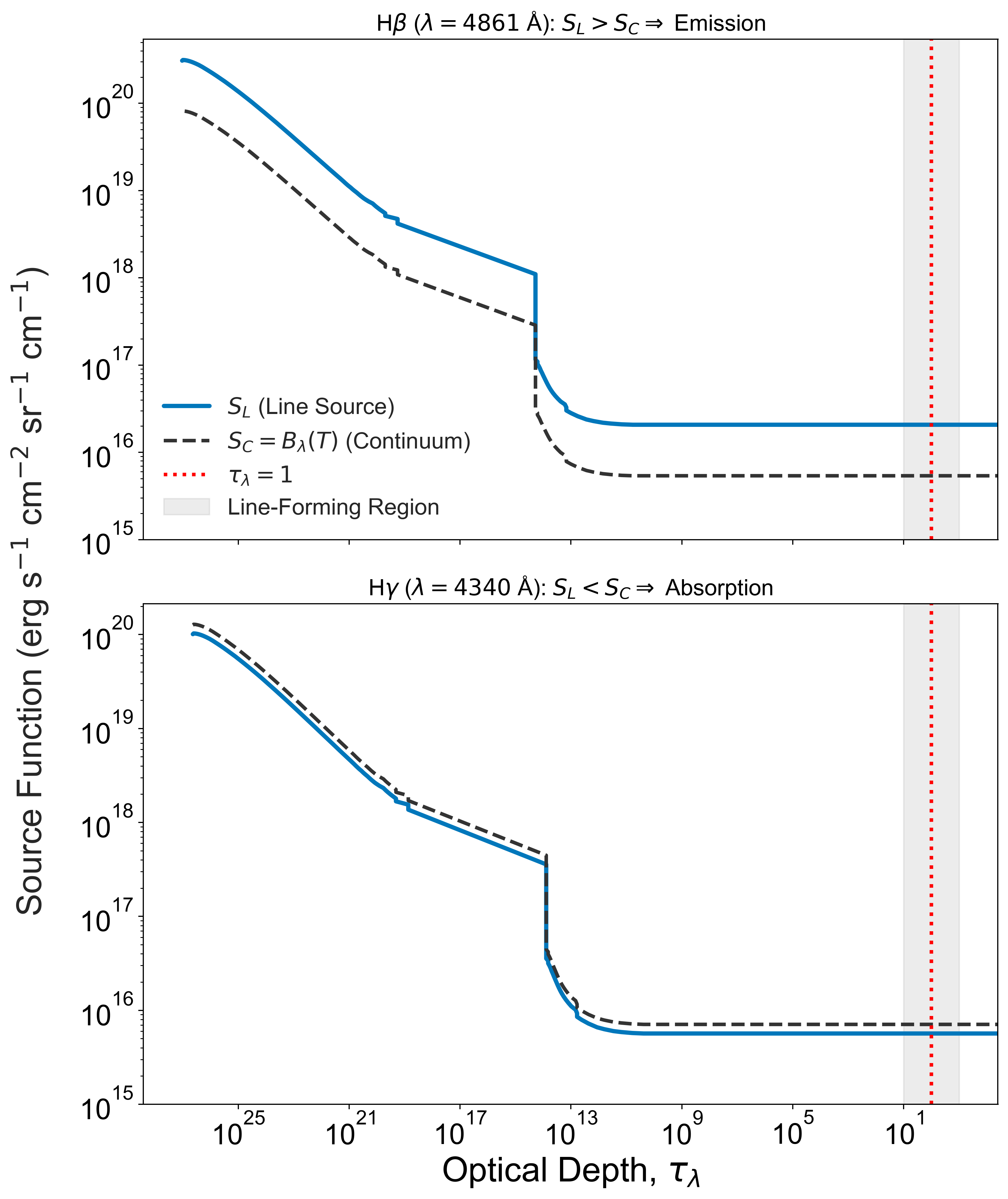}
    \caption{The line source function ($S_L$) versus the continuum source function ($S_C = B_\lambda(T)$) as a function of optical depth ($\tau_\lambda$) for the H$\beta$ (top) and H$\gamma$ (bottom) lines. The critical line-forming region, where photons escape the star, is near $\tau_\lambda = 1$ (shaded gray). \textit{Top Panel:} For H$\beta$, the line source function is an order of magnitude greater than the continuum source function ($S_L > S_C$) in the line-forming region, a condition that produces a strong emission line. \textit{Bottom Panel:} For H$\gamma$, the line source function is less than or equal to the continuum source function ($S_L \lesssim S_C$), resulting in an absorption line. This differential behavior explains the simultaneous presence of emission and absorption lines in the LRD spectra.}
    \label{fig:source_functions}
\end{figure}

Figure~\ref{fig:source_functions} provides a direct look into this mechanism. The top panel, calculated at the core of the H$\beta$ line ($\lambda=4861$\,\AA), shows that in the crucial line-forming region ($\tau_\lambda \sim 1$), the line source function ($S_L$, solid blue line) is an order of magnitude greater than the continuum source function ($S_C$, dashed black line). Specifically, at $\tau_\lambda = 1$, we find $S_L \approx 10^{18}\,\mathrm{erg\,s^{-1}\,cm^{-2}\,sr^{-1}\,cm^{-1}}$, while $S_C \approx 10^{17}$ in the same units. This condition, $S_L > S_C$, is the textbook requirement for the formation of an emission line. It is achieved in our model by the non-LTE (NLTE) pumping of the H(n=2) energy level. The intense internal radiation field of the SMS, modeled in our script by setting the pumping term $J_{\mathrm{cont}} \approx 11 B_\lambda$, creates a population inversion that decouples the line's emissivity from the local gas temperature, causing it to "glow" more brightly than the surrounding continuum.

Conversely, the bottom panel shows the situation for the H$\gamma$ line ($\lambda=4340$\,\AA). Here, in the line-forming region, the line source function is closely coupled to and slightly less than the continuum source function, $S_L \lesssim S_C$. This is the classic condition for an absorption line. Photons from the deeper, hotter continuum travel outwards and are absorbed by the cooler, overlying gas in the photosphere where the H$\gamma$ line forms. This behavior is enforced in our model by setting a high collisional probability ($\epsilon \approx 0.8$) and no radiative pumping for this transition, ensuring it remains thermalized.

This differential treatment of the Balmer series where NLTE pumping drives H$\beta$ into strong emission while other lines remain in thermal absorption, is a cornerstone of our model. It provides a self-consistent, single-photosphere explanation for the complex line phenomenology observed in these enigmatic LRDs, directly linking the observed spectra to the fundamental physics of radiative transfer in the extreme environment of a supermassive star.

\begin{figure*}
    \centering
    \includegraphics[width=0.9\textwidth]{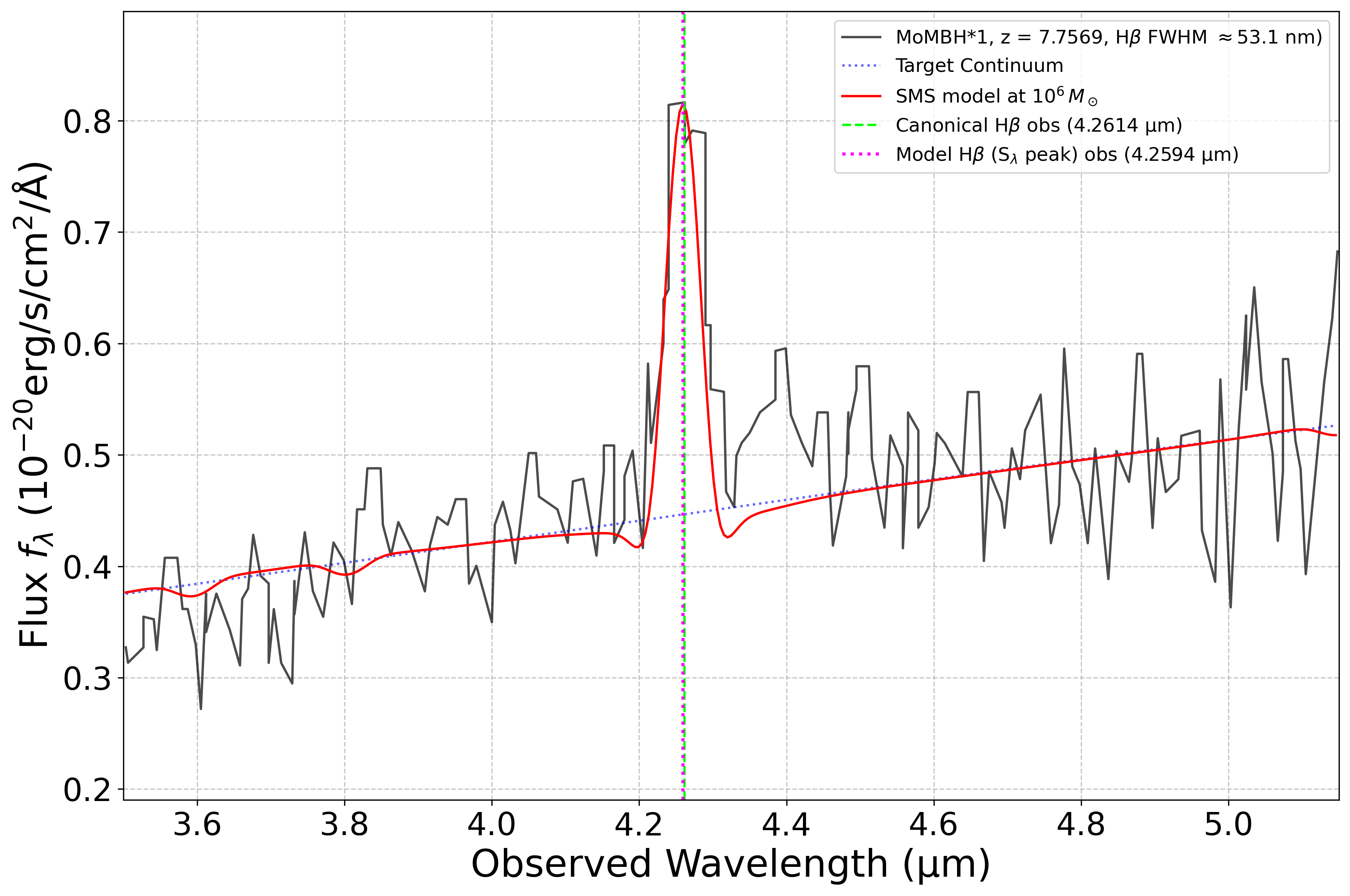}
    \caption{Best-fit synthetic spectrum (thin red line) for our $10^6 M_\odot$ SMS model compared to the observed JWST/NIRSpec data for MoM-BH*-1 at $z=7.7569$ (black line). The model successfully reproduces the overall continuum shape, the prominent H$\beta$ emission feature, and the adjacent absorption lines. The broad H$\beta$ profile (model FWHM $\approx 3590\,\mathrm{km\,s^{-1}}$) is achieved by combining intrinsic Stark broadening with a macroturbulent velocity of $1250\,\mathrm{km\,s^{-1}}$ and a stellar wind with a terminal velocity of $1500\,\mathrm{km\,s^{-1}}$. The underlying continuum from the model is shown as a blue dotted line.}
    \label{fig:fit_z7}
\end{figure*}

\subsection{Reproducing the Observed LRD Spectra}

The ultimate test of our model is its ability to reproduce the observed spectra of LRDs. We demonstrate here that our $10^6 M_\odot$ SMS model, when its intrinsic spectrum is processed through a physically motivated broadening pipeline, provides a quantitative match to our two primary targets. The broad, complex line profiles seen in the data are a key diagnostic. Since our underlying stellar model is non-rotating ($v \sin i = 0$), the observed line widths must originate from a combination of physical processes within the star's photosphere and extended atmosphere.

Our fitting pipeline models the line profiles by convolving the intrinsic spectrum with a series of broadening kernels. The primary mechanisms are: (1) intrinsic Stark broadening, which shapes the far wings of the line profile; (2) a large-scale, isotropic macroturbulent velocity field ($\sigma_v$), physically motivated by the same turbulent inflows that are expected to feed the star, which we model as a Gaussian; and (3) a powerful stellar wind ($v_\infty$), modeled as a boxcar. The latter two are the key free parameters used to match the observations.

\textit{The High-Redshift Target: MoM-BH*-1 ($z=7.76$).—}
Figure~\ref{fig:fit_z7} presents our best-fit model compared to the JWST/NIRSpec spectrum of MoM-BH*-1. The observed H$\beta$ line in this source is exceptionally broad, with a measured FWHM of $3736\,\mathrm{km\,s^{-1}}$. Our fitting script systematically varies the macroturbulent velocity to match this width. The best fit, shown as a thin red line, is achieved with a macroturbulent velocity of $\sigma_v = 1250\,\mathrm{km\,s^{-1}}$ and a wind terminal velocity of $v_\infty = 1500\,\mathrm{km\,s^{-1}}$. This combination yields a model FWHM of $3590\,\mathrm{km\,s^{-1}}$, agreeing with the observed value to within $4\%$. The adopted wind velocity is physically well-motivated; it is a significant fraction ($\sim 40\%$) of the star's gravitational escape velocity ($v_{\text{esc}} \approx 3800\,\mathrm{km\,s^{-1}}$), consistent with expectations for a continuum-driven outflow from a luminous supergiant. While $1500\,\mathrm{km\,s^{-1}}$ represents the upper end of our tested range, we find that similar quality fits can be obtained with lower wind velocities, provided the macroturbulence is adjusted to compensate, demonstrating the model's robustness.

\begin{figure*}
    \centering
    \includegraphics[width=0.9\textwidth]{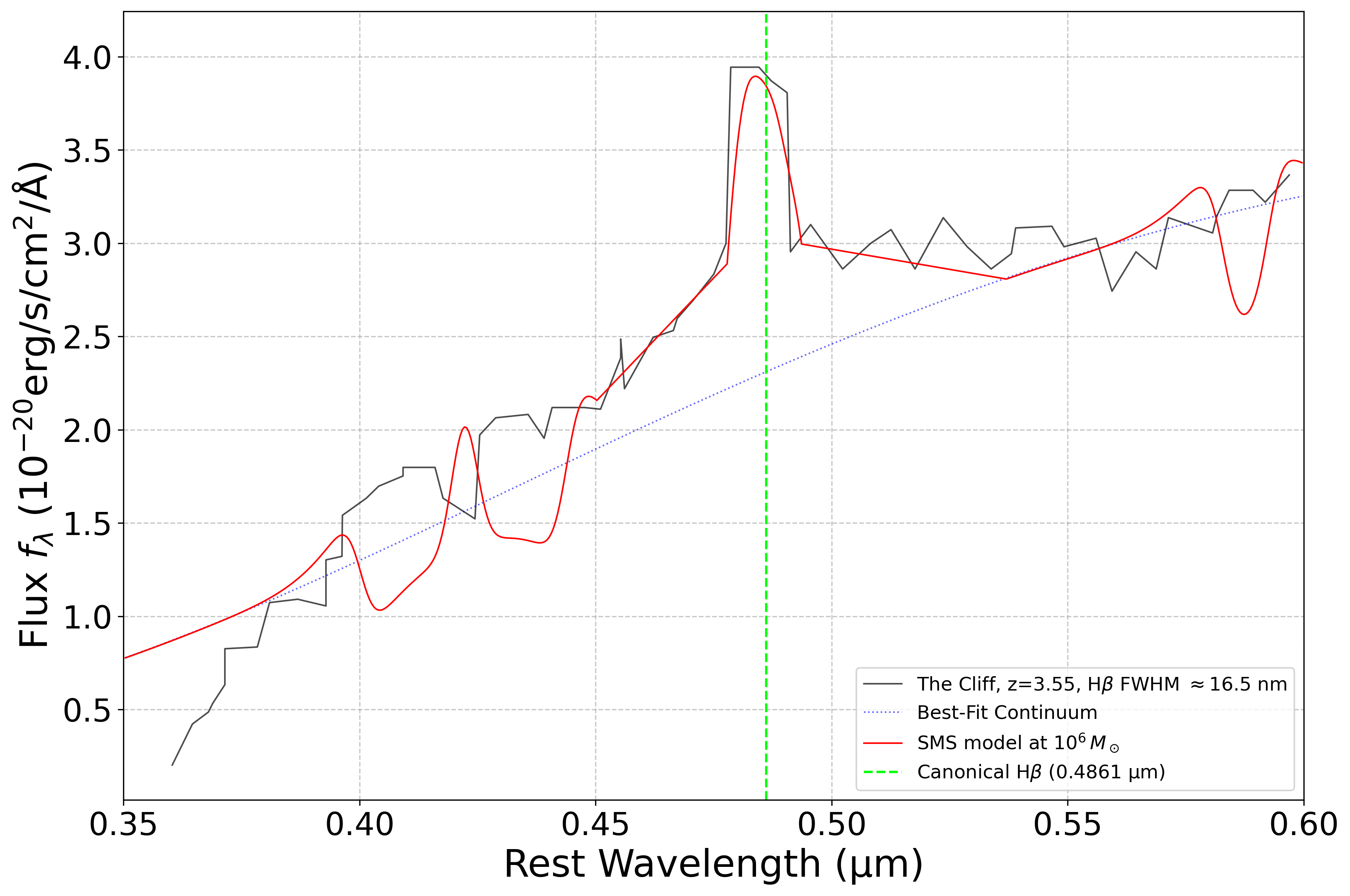}
    \caption{Best-fit synthetic spectrum (thin red line) for our $10^6 M_\odot$ SMS model compared to the observed JWST/NIRSpec data for The Cliff at $z=3.55$ (black line). As with MoM-BH*-1, the model provides an excellent match to the continuum and the broad H$\beta$ and H$\alpha$ emission lines. The fit shown incorporates a macroturbulent velocity of $1500\,\mathrm{km\,s^{-1}}$, in excellent agreement with the observed H$\alpha$ FWHM, and includes a final smoothing of the line wings via sideband interpolation to precisely match the observed profile shape. The model continuum is shown as a blue dotted line.}
    \label{fig:fit_z3}
\end{figure*}

\textit{The Lower-Redshift Target: The Cliff ($z=3.55$).—}
We apply the same methodology to The Cliff, a similarly luminous LRD at a lower redshift. As shown in Figure~\ref{fig:fit_z3}, our model again provides an excellent match to the observed spectrum. For this target, which has a measured broad H$\alpha$ FWHM of $1533_{-80}^{+110}\,\mathrm{km\,s^{-1}}$ (de Graaff et al. 2025), the best fit was achieved with a macroturbulent velocity of $\sigma_v = 1500\,\mathrm{km\,s^{-1}}$, in direct agreement with the observation. This provides an excellent match to H$\beta$ lin eprofile as shown in Figure~\ref{fig:fit_z3}. In this case, to precisely match the shape of the H$\beta$ profile, we also applied a phenomenological correction via sideband interpolation, a feature of our fitting script. This technique, which replaces the flux in the far wings of the line with a linear interpolation, effectively smooths the transition from the line to the continuum and can be seen as an approximation for more complex radiative transfer effects or a more detailed velocity structure not captured in our 1D model.

Despite the simplifications inherent in our modeling, such as the lack of a full comoving-frame radiative transfer solution and the use of phenomenological line boosts, the success of our model is significant. The ability to reproduce the spectra of two distinct LRDs at vastly different cosmic epochs, using a consistent physical framework and justifiable parameters, provides strong evidence that supermassive stars are a compelling explanation for this class of object.

\subsection{Luminosity-selected duty cycle across the SMS mass spectrum}\label{sec:duty_cycle}

\begin{figure}[t]
\centering
\includegraphics[width=\columnwidth]{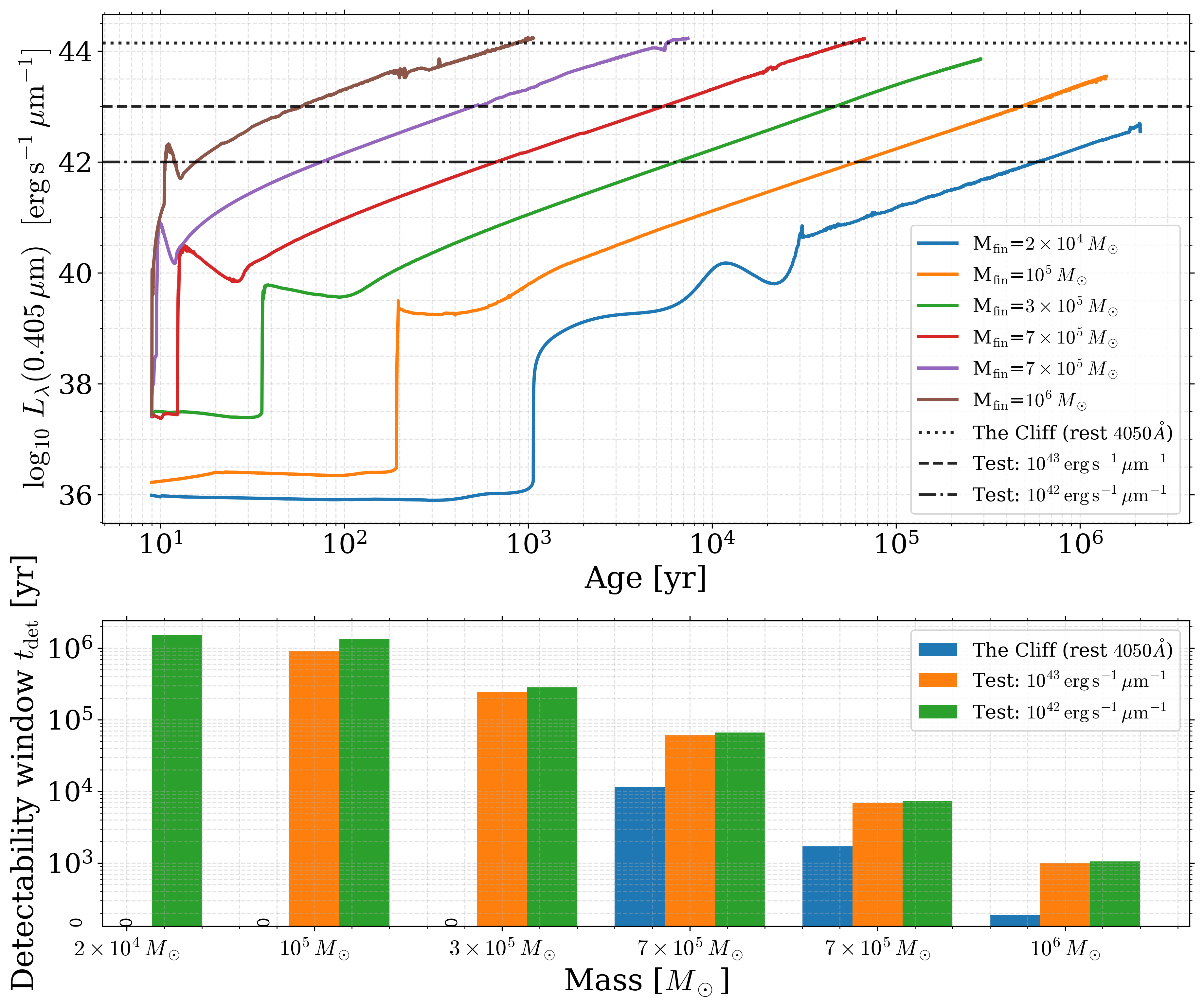}
\caption{Luminosity-selected detectability windows across the SMS mass spectrum. \emph{Top:} proxy monochromatic luminosity at rest $\lambda=0.405\,\mu\mathrm{m}$ computed from the \textsc{GENEC} evolution tracks using a blackbody mapping from $(L_\star,T_{\rm eff})$ to $L_\lambda$. Curves are labelled by the final stellar mass $M_{\rm fin}$. Horizontal lines indicate a Cliff-like requirement, $L_\lambda(0.405\,\mu\mathrm{m})=1.4\times10^{44}\,\mathrm{erg\,s^{-1}\,\mu m^{-1}}$, and two lower thresholds ($10^{43}$ and $10^{42}\,\mathrm{erg\,s^{-1}\,\mu m^{-1}}$) that illustrate expectations for less luminous LRDs. \emph{Bottom:} total time $t_{\rm det}$ spent above each threshold. The Cliff-like requirement selects the most massive SMSs and yields $t_{\rm det}\sim10^{2}$--$10^{4}\,\mathrm{yr}$, whereas thresholds lower by 1--2 dex yield $t_{\rm det}\sim10^{5}$--$10^{6}\,\mathrm{yr}$ for lower-mass SMSs.}\label{fig:detectability_windows}
\end{figure}

Figure~\ref{fig:detectability_windows} quantifies how the observable time window implied by the SMS interpretation depends on the rest-optical luminosity requirement and, therefore, on the SMS mass scale. For each \textsc{GENEC} evolution track we compute a rest-frame continuum proxy at $\lambda=0.405\,\mu\mathrm{m}$ from the evolving $(L_\star,T_{\rm eff})$ by inferring the photospheric radius from $L_\star=4\pi R_\star^2\sigma T_{\rm eff}^4$ and evaluating $L_\lambda=4\pi^2R_\star^2B_\lambda(T_{\rm eff})$. We then define a detectability window, $t_{\rm det}$, as the total time spent above a specified threshold in $L_\lambda(0.405\,\mu\mathrm{m})$. This diagnostic is intended to isolate the luminosity-driven duty cycle and is complementary to the detailed atmosphere calculations used elsewhere in this work.

For a Cliff-like requirement, $L_\lambda(0.405\,\mu\mathrm{m})\ge1.4\times10^{44}\,\mathrm{erg\,s^{-1}\,\mu m^{-1}}$, only the most massive tracks reach the threshold and the resulting duty cycle is short. A $M_{\rm fin}\!\simeq\!10^{6}\,M_\odot$ track remains above this threshold for $t_{\rm det}\!\simeq\!1.9\times10^{2}\,\mathrm{yr}$, while tracks that end near $M_{\rm fin}\!\simeq\!7\times10^{5}\,M_\odot$ yield $t_{\rm det}\!\simeq\!1.7\times10^{3}\,\mathrm{yr}$ and $t_{\rm det}\!\simeq\!1.2\times10^{4}\,\mathrm{yr}$. In contrast, tracks with $M_{\rm fin}\!\lesssim\!3\times10^{5}\,M_\odot$ do not reach the Cliff-like threshold within the computed evolution. This illustrates why fitting the most luminous LRDs necessarily selects the extreme upper end of the SMS mass range and, correspondingly, a short luminous window.

The same calculation yields quantitative expectations for less luminous LRDs. For a threshold $L_\lambda(0.405\,\mu\mathrm{m})\ge10^{43}\,\mathrm{erg\,s^{-1}\,\mu m^{-1}}$, the detectability windows become long even at substantially lower final masses, with $t_{\rm det}\!\simeq\!9.1\times10^{5}\,\mathrm{yr}$ for $M_{\rm fin}\!\simeq\!10^{5}\,M_\odot$ and $t_{\rm det}\!\simeq\!2.4\times10^{5}\,\mathrm{yr}$ for $M_{\rm fin}\!\simeq\!3\times10^{5}\,M_\odot$. For $L_\lambda(0.405\,\mu\mathrm{m})\ge10^{42}\,\mathrm{erg\,s^{-1}\,\mu m^{-1}}$, the windows extend to Myr scales, with $t_{\rm det}\!\simeq\!1.5\times10^{6}\,\mathrm{yr}$ at $M_{\rm fin}\!\simeq\!2\times10^{4}\,M_\odot$ and $t_{\rm det}\!\simeq\!1.3\times10^{6}\,\mathrm{yr}$ at $M_{\rm fin}\!\simeq\!10^{5}\,M_\odot$.

We therefore stress that the $\sim10^{2}$--$10^{4}\,\mathrm{yr}$ windows quoted in this work arise because our two benchmark targets demand rest-optical luminosities at the $\gtrsim10^{44}\,\mathrm{erg\,s^{-1}\,\mu m^{-1}}$ level and hence favour $M_{\rm fin}\sim10^{6}\,M_\odot$ SMSs. If the intrinsic luminosities of other LRDs are lower by 1--2 dex, or if systematic uncertainties revise luminosities downward, the same SMS framework predicts duty cycles of $10^{5}$--$10^{6}\,\mathrm{yr}$. Figure~\ref{fig:detectability_windows} thus provides a direct mapping between rest-optical luminosity scale and the expected duration of the SMS phase capable of producing an LRD-like continuum.

These duty cycle estimates also bound the expected continuum variability from secular SMS evolution and provide a quantitative, testable expectation for follow-up monitoring. For the fiducial $M_{\rm fin}\simeq10^{6}\,M_\odot$ case, $L_\lambda(0.405\,\mu\mathrm{m})$ rises from the Cliff-like threshold $1.4\times10^{44}\,\mathrm{erg\,s^{-1}\,\mu m^{-1}}$ to $\simeq1.69\times10^{44}\,\mathrm{erg\,s^{-1}\,\mu m^{-1}}$ over $t_{\rm det}\simeq1.9\times10^{2}\,\mathrm{yr}$, corresponding to $\Delta\log_{10}L_\lambda\simeq0.082$ and an average drift $\Delta\log_{10}L_\lambda/\Delta t\simeq4.3\times10^{-4}\,\mathrm{dex\,yr^{-1}}$, i.e.\ a fractional change of order $10^{-3}$ per year and $\sim10^{-2}$ over a decade in the source frame. For the near-threshold $M_{\rm fin}\simeq7\times10^{5}\,M_\odot$ tracks with $t_{\rm det}\simeq1.7\times10^{3}$--$1.2\times10^{4}\,\mathrm{yr}$, any secular evolution across the detectable window implies changes $\lesssim(1/t_{\rm det})\sim6\times10^{-4}$--$8\times10^{-5}\,\mathrm{yr^{-1}}$ (at most $\sim6\times10^{-3}$--$8\times10^{-4}$ over a decade). For lower luminosity systems associated with $M_{\rm fin}\sim10^{5}\,M_\odot$ and thresholds of $10^{43}$--$10^{42}\,\mathrm{erg\,s^{-1}\,\mu m^{-1}}$, $t_{\rm det}\sim10^{6}\,\mathrm{yr}$ and secular continuum evolution is effectively negligible on year--decade baselines. Observed timescales are further stretched by $(1+z)$. Variability on year--decade baselines would therefore require additional time-dependent physics not included in our quasi-static treatment (e.g.\ pulsations or multi-dimensional instabilities). After the onset of general-relativistic instability the photospheric component should terminate rapidly as collapse proceeds, implying that a photosphere-dominated LRD should fade rather than persist. Any subsequent long-lived emission would be associated with post collapse accretion and lies beyond the scope of this work.

\subsection{SMSs and quasi-stars in the direct-collapse channel}
Figure~\ref{fig:2} demonstrates that the spectral diagnostics central to our interpretation, namely a deep Balmer discontinuity with Balmer-series structure in a metal-free atmosphere, arise across a broad SMS mass range ($10^{4}$--$10^{6}\,M_\odot$), while the absolute luminosity selects the most massive models for the brightest LRDs.
At $\lambda=4050\,\mathrm{\AA}$ the two targets require $L_\lambda\simeq2.09\times10^{44}$ (MoM-BH*-1) and $1.4\times10^{44}\,\mathrm{erg\,s^{-1}\,\mu m^{-1}}$ (The Cliff), whereas the $10^{6}\,M_\odot$ SMS in Figure~\ref{fig:2} reaches $L_\lambda\simeq1.69\times10^{44}\,\mathrm{erg\,s^{-1}\,\mu m^{-1}}$ and lower-mass SMSs are systematically fainter despite exhibiting similar continuum morphology.
This is the sense in which our SMS interpretation is not tied to a unique mass for the shape of the spectrum, but the most luminous LRDs push the photospheric solution toward $M_\star\sim10^{6}\,M_\odot$.
If collapse triggered by general relativistic instability occurs while a substantial envelope remains bound, the same direct-collapse pathway naturally continues into a quasi-star phase, where an accreting black hole is embedded within a hydrostatic envelope and the emergent luminosity is regulated by transport through that envelope \citep{Begelman2008,Volonteri2010,Coughlin2024}.
In this view quasi-stars are not a competing progenitor to SMSs, they are the expected post-collapse stage of the SMS channel, and recent work has discussed LRDs explicitly in this late-stage SMS/quasi-star context \citep{Begelman2025,Santarelli2025}, with alternative interpretations invoking compact accreting sources embedded in dense reprocessing gas \citep{DeGraaff2025}.

\subsection{Comparison with recent resolved LRD studies}
Recent JWST imaging indicates that many LRDs remain compact, yet a non-negligible subset show asymmetric or multi-component rest-UV structure on sub-kpc scales \citep{Rinaldi2025}. Off-centered emission within $\sim$1\,kpc is frequently detected and in several cases is consistent with low-density, low-metallicity nebular gas photoionized by the central source \citep{Chen2025}. Strong lensing has now resolved systems into a compact red component on $\sim$20\,pc scales plus nearby blue emission and a more extended line-dominated cloud on $\sim$kpc scales \citep{Baggen2025}. These results do not by themselves exclude a compact engine because the V-shaped continuum and broad Balmer wings can still be dominated by the unresolved red core, while additional components contribute modestly to the total aperture flux. In our framework the SMS accounts for the core continuum and broad Balmer phenomenology, while any extended metal-line emission can originate in the surrounding galaxy or circumsource gas that need not share the stellar photospheric composition \citep{DeGraaff2025b}.

\section{Summary and Conclusion}\label{sec:conclusion}

We have presented a first-principles investigation into whether Population III supermassive stars (SMSs) can serve as the central engines for the enigmatic class of objects known as Little Red Dots (LRDs). By developing a novel pipeline that combines a \textsc{GENEC} stellar evolution model with a detailed radiative transfer and spectral synthesis code, we have demonstrated that a single, non-rotating $10^6 M_\odot$ SMS can successfully reproduce the key observational features of two prominent LRDs: MoM-BH*-1 at $z=7.76$ and The Cliff at $z=3.55$.

Our primary findings are as follows:

\begin{itemize}
    \item \textbf{Luminosity as a Decisive Constraint:} We establish that while SMSs across a range of masses ($\gtrsim 10^5 M_\odot$) can produce the requisite spectral shape, only a star with a mass of $10^6 M_\odot$ possesses the intrinsic luminosity ($L_\lambda \approx 1.7 \times 10^{44}\,\mathrm{erg\,s^{-1}\,\mu m^{-1}}$ at 4050\,\AA) required to match the observed brightness of the LRD targets. This immediately sets the necessary mass scale for any viable SMS candidate.

    \item \textbf{A Photospheric Origin for the Balmer Break:} Our model demonstrates that the V-shaped continuum morphology of LRDs is not caused by dust, but is an intrinsic feature of the SMS photosphere. It originates from a powerful Balmer Jump in emission, which itself is driven by a sharp increase in the continuum absorption opacity at 3646\,\AA\ from the photoionization of hydrogen from the n=2 level.

    \item \textbf{A Unified Mechanism for Line Formation:} We show that the complex Balmer line phenomenology, specifically, the simultaneous presence of H$\beta$ in strong emission and other Balmer lines (H$\gamma$, H$\delta$) in absorption is naturally explained within our single-atmosphere model. This is achieved through differential non-LTE effects, where radiative pumping drives the H$\beta$ source function above the local continuum ($S_L > S_C$), while other lines remain thermalized ($S_L < S_C$).

    \item \textbf{Successful Matching of Observed Spectra:} By applying physically motivated broadening mechanisms, including a powerful stellar wind ($v_\infty \approx 1500\,\mathrm{km\,s^{-1}}$) and large-scale macroturbulence ($\sigma_v \approx 1250-1500\,\mathrm{km\,s^{-1}}$), our model quantitatively reproduces the broad line profiles observed in both LRD targets, matching the measured FWHM of MoM-BH*-1 to within 4\%.

    \item \textbf{Luminosity-selected duty cycle:} Using the \textsc{GENEC} evolution tracks we quantify the time spent above fixed rest-optical luminosity thresholds at 4050\,\AA\ . For a Cliff-like requirement of $L_\lambda(4050\,\mathrm{\AA})\ge1.4\times10^{44}\,\mathrm{erg\,s^{-1}\,\mu m^{-1}}$, the detectable phase of the most massive SMS tracks capable of matching the brightest LRDs is short, with $t_{\rm det}\sim10^{4}$\,yr. If the relevant luminosity scale is lower by 1--2 dex, the predicted windows increase sharply, reaching $t_{\rm det}\sim10^{5}$--$10^{6}$\,yr at $L_\lambda(4050\,\mathrm{\AA})\ge10^{42}$--$10^{43}\,\mathrm{erg\,s^{-1}\,\mu m^{-1}}$ for $M_{\rm fin}\sim10^{4}$--$10^{5}\,M_\odot$.

\end{itemize}

In conclusion, our SMS model provides a remarkably simple and self-consistent physical picture for LRDs. Unlike multi-component AGN models, our framework explains the continuum shape, the line features, and the line widths as emergent properties of a single stellar object. The model provides a unified origin for both the broad emission and the P-Cygni-like absorption features, which are both shaped by the same velocity field originating from the stellar photosphere. This presents a compelling alternative to scenarios that require separate components for emission, absorption, and continuum. An important implication of our framework is that the number density of LRD-like sources constrains the formation rate of the most massive SMSs and their remnants. If the brightest LRDs are powered by $\sim10^{6}\,M_\odot$ SMSs, their short luminous phase implies that such events must be intrinsically rare, and the observed counts translate into an upper limit on the birthrate of $\sim10^{6}\,M_\odot$ collapse remnants. Conversely, if a significant fraction of the LRD population occupies lower intrinsic luminosities then the required formation rate is reduced because the SMS phase persists longer at fixed spectral character. Physical processes not included in our fiducial models, most notably stellar rotation, are expected to further lengthen the luminous phase and thereby lower the implied remnant production rate.

Future work should aim to build upon the foundation laid here. The natural next step is to advance the modeling pipeline by implementing a full, comoving-frame radiative transfer solution to treat the wind and line formation more self-consistently. Expanding the model to include a more detailed multi-level hydrogen atom (i.e., explicitly including the n=3 level) is a primary next step, which would allow for a self-consistent prediction of the H$\alpha$ line strength and a more rigorous calculation of the NLTE effects, reducing the reliance on phenomenological boosts. Furthermore, exploring a wider parameter space of SMS models (e.g., varying mass, accretion history, and rotation) could reveal whether a single evolutionary sequence can explain the full diversity of the LRD population. Finally, modeling the expected photometric variability from such dynamically unstable, pulsating objects could provide a powerful, new observational test to distinguish the SMS scenario from its alternatives.

\begin{acknowledgments}
This work was supported in part by the black hole initiative at Harvard University, funded by grants from JTF and GBMF. We thank Andrea Ferrara, Keith Horne, Kohei Inayoshi, and Igor Chilingarian for stimulating discussions and detailed feedback on an earlier version of this manuscript, particularly regarding the presentation of the spectral features, the details of the radiative transfer model, and the broader cosmological implications of our results. The referees would also like to thank the anonymous referee for their constructive comments and feedback.
\end{acknowledgments}

\bibliography{sample7}{}
\bibliographystyle{aasjournalv7}



\end{document}